\documentclass{article}
\usepackage{hiph-art}
\volnumber{19} \issuenumber{1} \edyear{2004}                             
\frompage{000} \topage{000}                                              
\recrevdate{\today}                                                      
\def\rightmark{
What can we learn from hydrodynamic analysis at RHIC?
}
\def\leftmark{T. Hirano}
 
\title{
What can we learn from hydrodynamic analysis at RHIC?
} 
\authors{ 
{Tetsufumi Hirano$^{1,a}$ %
\index{Hirano, T.} 
}\\[2.812mm]
{\normalsize
\hspace*{-8pt}$^1$ Department of Physics,
	Columbia University\\ 
New York, NY 10027, USA\\[0.2ex] 
}}
 
\abstract{
We can establish a new picture, 
the perfect fluid sQGP core and
the dissipative hadronic corona,
of the space-time evolution of
produced matter in relativistic heavy ion collisions
at RHIC.
It is also shown that the picture works well
also in the forward rapidity region through an analysis
based on a new class of the hydro-kinetic model
and that this is a manifestation of rapid increase
of entropy density in the vicinity of QCD critical
temperature, namely deconfinement.
}

 
\makeindex
\begin{document}
 
\maketitle

\section{Introduction}\label{intro}
Recently, physicists at Brookhaven National Laboratory
made an announcement 
in the American Physical Society
annual meeting at Florida, USA, April 18, 2005,
that ``RHIC serves the perfect liquid" \cite{BNL}.
The agreement of hydrodynamic predictions
\cite{Huovinen:2003fa}
of integrated
and differential elliptic flow and radial flow patterns
with Au+Au data at RHIC energies
\cite{Ackermann:2000tr,Adcox:2002ms,Back:2002gz,Ito} 
is one of the main lines of the announcement.
We first study the sensitivity of this conclusion
to different hydrodynamic assumptions 
in the hadron phase.
It is found that an assumption of chemical equilibrium
with neglecting viscosity in the hadron phase
in hydrodynamic simulations causes accidental reproduction
of transverse momentum spectra and 
differential elliptic flow data
in a way that chemical equilibrium imitates a sort of
dissipation.
From a systematic comparison of hydrodynamic
results with the experimental data, dissipative effects
are found to be mandatory in the hadron phase.
Therefore, what is discovered at RHIC is not only
the perfect fluidity of the strongly coupled
quark gluon plasma (sQGP) core
but also its dissipative hadronic corona
surrounding the core of the sQGP.
Along the lines of these studies,
we develop a hybrid dynamical model
in which a \textit{fully three-dimensional}
hydrodynamic description of the QGP phase
is followed by a kinetic description of the hadron phase 
\cite{nonaka}.
We will show rapidity dependence of elliptic flow from this hybrid model
supports the above picture in Sec.~3. 
Finally, we will argue in Sec.~4 
that this picture is a manifestation of
deconfinement transition, namely,
a rapid increase of entropy density
in the vicinity of the QCD critical temperature
as lattice QCD simulations have been predicted \cite{Karsch}.

\section{SQGP CORE AND DISSIPATIVE HADRONIC CORONA}\label{sqgpcore}  

A perfect fluid in the QGP phase is assumed 
in most hydrodynamic simulations.
While one can find various assumptions 
in the hadron phase \cite{Huovinen:2003fa},
e.g.
(1) an ideal fluid in chemical equilibrium (CE),
(2) an ideal fluid in which particle ratios are fixed
(or in partial chemical equilibrium, PCE), or
(3) a non-equilibrium resonance gas via hadronic cascade models (HC).
Hydrodynamic results are compared with the current
differential elliptic flow data, $v_2(p_T)$,
in Fig.~20 in Ref.~\cite{Adcox:2004mh}
with putting an emphasis on the difference of
these assumptions in the hadron phase.
The classes CE and HC reproduce
the data for pions and protons well.
This is considered to be one of the evidence of success of
an ideal QGP fluid at RHIC.
Contrary to its success, results from the second class, PCE, 
deviate from the other hydrodynamic results and
experimental data though the class PCE as an ideal fluid 
describes chemical compositions
of the hadronic matter
in a more realistic way than the class CE does.
In order to claim the discovery of perfect fluidity
from the agreement of hydrodynamic results 
with $v_2(p_T)$ data,
we need to understand the difference among hydrodynamic
results and the deviation from data within the PCE assumption.
The difference of assumptions in the hadron phases 
reduces to the difference of 
the final slope of differential
elliptic flow $dv_2(p_T)/dp_T$.
$v_2$ is roughly proportional to $p_T$
in low $p_T$ region for pions.
In such a case, the slope of $v_2(p_T)$
can be approximated by
$v_2/\langle p_T \rangle$
since one can easily show 
$dv_2(p_T)/dp_T = v_2/\langle p_T \rangle$ when $v_2(p_T)$ is
\textit{exactly} proportional to $p_T$.
Integrated $v_2$ is generated in the early
stage of collisions. Whereas \textit{differential}
$v_2$ can be sensitive to the late hadronic stage
since $dv_2(p_T)/dp_T \approx v_2/\langle p_T \rangle$
indicates interplay between elliptic flow
(integrated $v_2$) 
and radial flow (mean transverse momentum $\langle p_T \rangle$).

\begin{figure}[htb]
\insertplot{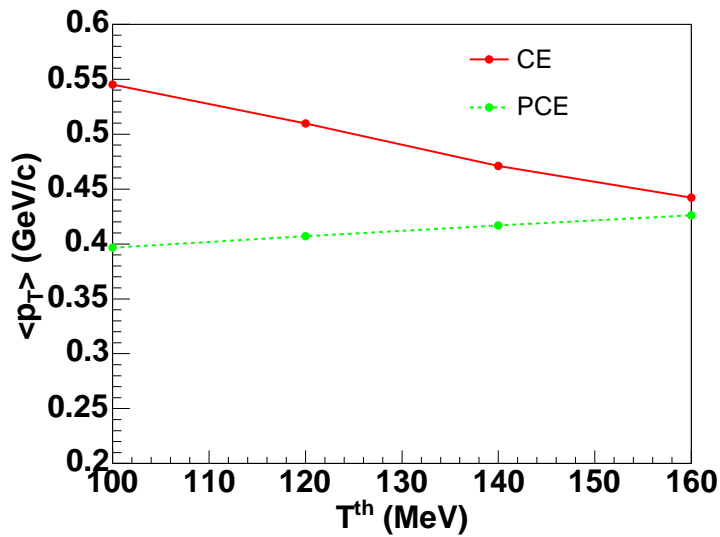}
\caption[]{Mean transverse momentum for
pions as a function of thermal
freezeout temperature at $b=5$ fm in $\sqrt{s_{NN}}=200$ GeV
Au+Au collisions.
Solid (dashed) line shows a result with
an assumption of an ideal, chemical equilibrium (chemically frozen)
hadronic fluid.}
\label{fig:meanpt}
\end{figure}

\begin{figure}[htb]
\insertplot{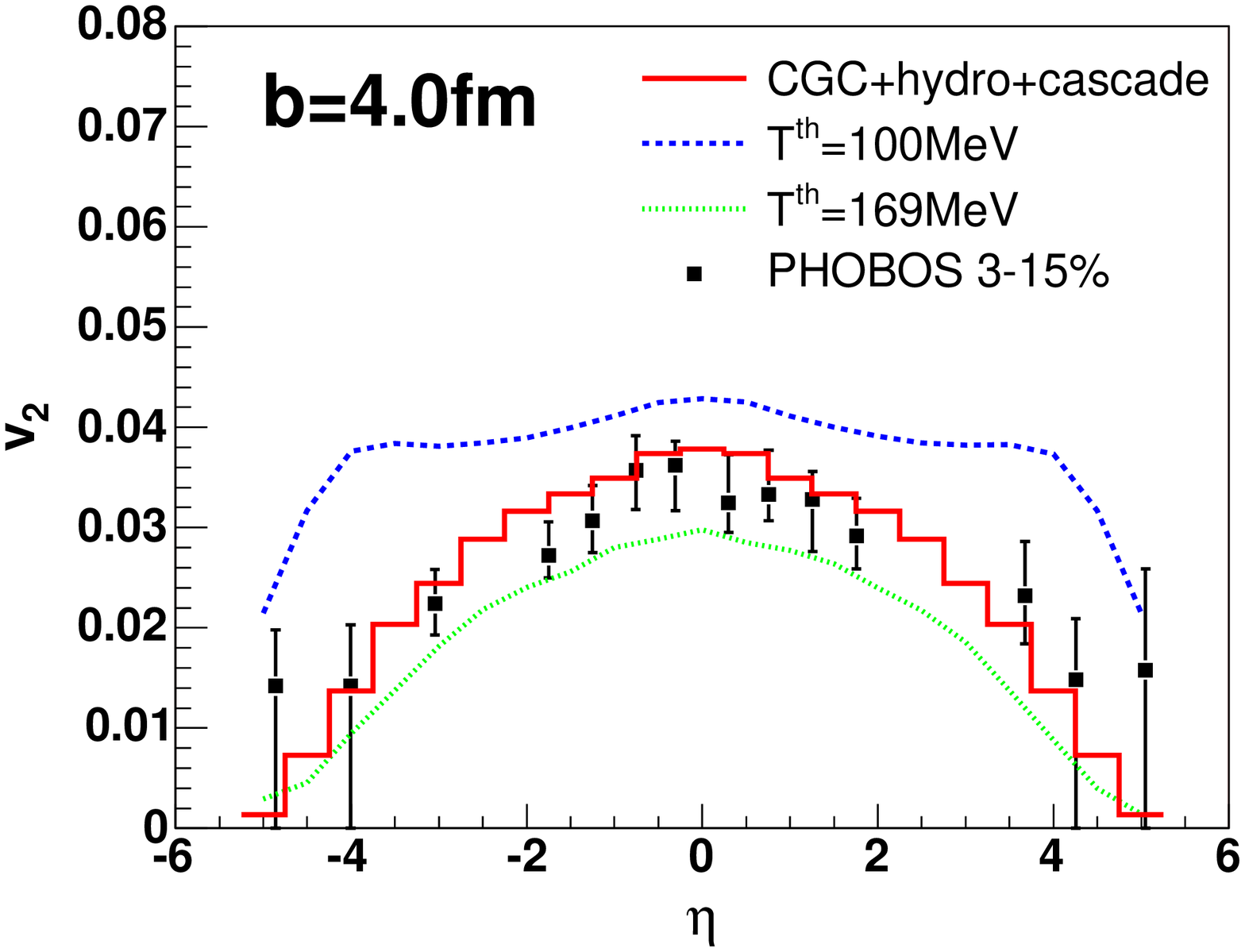}
\caption[]{Pseudorapidity dependence of elliptic flow
from hydro and hydro+hadronic cascade models are
compared with data \cite{Back:2002gz}.
Dashed and dotted lines are results from ideal hydrodynamics
with $T^{\mathrm{th}} = 100$ and 169 MeV respectively.
Solid line is a result from a hydro+cascade hybrid model.
}
\label{fig:v2eta}
\end{figure}
In Fig.~\ref{fig:meanpt},
thermal freezeout temperature $T^{\mathrm{th}}$ dependence of
$\langle p_T \rangle$ for pions including
contribution from resonance decays
are shown from
hydrodynamic simulations \cite{hg}.
Here the impact parameter $b=5$ fm
and collision energy $\sqrt{s_{NN}}=200$ GeV
are assumed in the hydrodynamic simulations.
$\langle p_T \rangle$ for pions 
in the chemically frozen hadronic fluid
decreases with decreasing $T^{\mathrm{th}}$
(increasing proper time $\tau$).
This is due to longitudinal $pdV$ work done by fluid elements.
Whereas $\langle p_T \rangle$ in the chemical equilibrium case
increases during expansion.
This counterintuitive result appears due to the assumptions
of entropy conservation and chemical equilibrium.
The total number of particles in a fluid element
decreases with proper time $\tau$ due to mass effects in the
hadron phase. Mass of particles
contributes to the entropy density significantly in low temperature and
the proportionality between the entropy density and the number density
violates in the temperature range under consideration.
Then the total energy that a fluid element possesses initially
is partly used for $pdV$ work due to expansion
as well as the case for chemically frozen fluids.
As a fluid element expands,
the remaining energy
is distributed among the smaller number
of particles in chemical equilibrium.
These are the reasons why the different behavior 
of $\langle p_T \rangle$ appears
according to the assumption of chemical equilibrium/freezeout.
Note that increase of $\langle p_T \rangle$
results in a ``reheating" behavior of a fluid
element:
Temperature in the chemical equilibrium hadronic fluid
drops more slowly than that in the chemically frozen fluid does
as if there exists a sort of dissipation  \cite{Hirano:2002ds}.
Under the chemical equilibrium assumption
in ideal hydrodynamic simulations,
increasing $\langle p_T \rangle$ is
commonly utilized so far to fix  
$T^{\mathrm{th}}$
by fitting $p_T$ slope.
However, this is attained only
by neglecting data of particle ratio.
If particle ratios are fixed
properly in hydrodynamic simulations
to reproduce the data,
$p_T$ slopes, especially for protons, are hardly reproduced
\cite{hiranonara} due to a lack of radial flow.
The same is true for differential elliptic flow:
$dv_2(p_T)/dp_T \approx v_2/\langle p_T \rangle$
is reproduced by canceling increasing behaviors of 
both $v_2$ and $\langle p_T \rangle$
under chemical equilibrium assumption \cite{hg}.
Agreement of the results from the CE model
with $p_T$ spectra and $v_2(p_T)$ data
is reached only when one neglects particle ratio
and the system keeps chemical equilibrium until kinetic
freezeout.
So, among three classes mentioned above,
the HC model turns out to be the only
model which is able to reproduce
particle ratio, $p_T$ spectra and $v_2(p_T)$ data at once
in a proper way.
Therefore a picture of 
the dissipative hadronic corona
together with
the perfect fluid sQGP core,
which is properly implemented in the HC model,
is consistent with these
experimental data observed at RHIC.

\section{3D HYDRO AND HADRONIC CASCADE MODEL}\label{hydrocascade}

Establishment of a dynamical model
for a detailed description of the whole space-time evolution 
is one of the main goals
in the physics of relativistic heavy ion collisions.
Dynamical models for description
of the whole space-time evolution
will be needed to draw the
transport properties
of the QGP from the experimental data.
A dynamical framework
has been developed \cite{hiranonara} to describe
three important aspects of relativistic heavy ion collisions,
namely colour glass condensate (CGC) for collisions of two
nuclei, ideal 3D hydrodynamics for space-time
evolution of locally thermalised matter, and jet quenching
for high $p_T$ non-thermalised matter.
According to the discussion
in the previous section,
we incorporate a hadronic cascade model JAM \cite{jam}
into our previous framework,
the ``CGC+hydro+jet" model \cite{hiranonara},
to establish a more realistic description of space-time evolution
as discussed in the previous section.

\begin{figure}[tb]
\insertplot{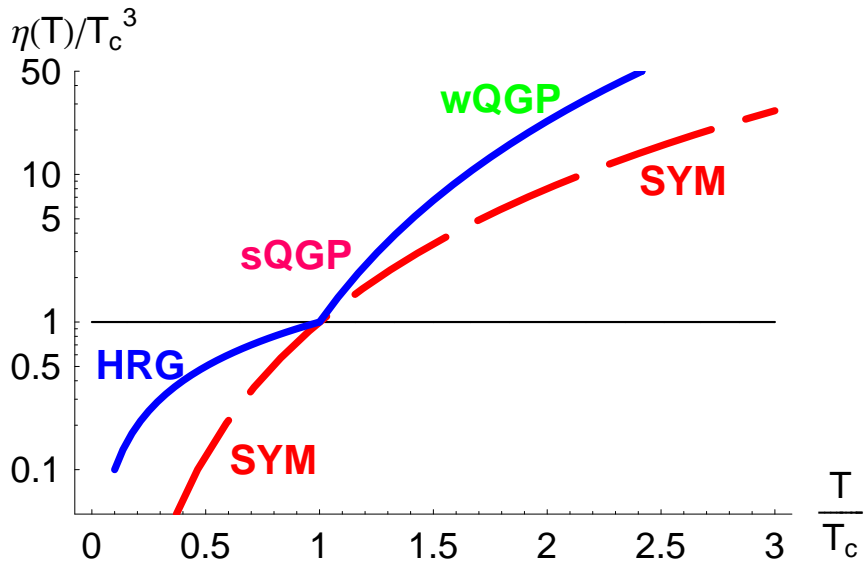}
\caption[]{Illustration of the approximately monotonic increase of
absolute value of the shear viscosity with temperature.
SYM, HRG, and wQGP represent, respectively, supersymmetric Yang-Mills model,
hadronic resonance gas, and weakly coupled QGP.
}
\label{fig:eta}
\end{figure}
Figure 2 shows pseudorapidity dependences of $v_2$
from this hybrid model 
and ideal 3D hydrodynamics with $T^{\mathrm{th}}=100$ and 169 MeV.
Here critical temperature
and chemical freezeout temperature
are taken as being 
$T_c = T^{\mathrm{ch}}=170$ MeV
in the hydrodynamic model.
In the hybrid model,
in which hydrodynamic description of
the sQGP ideal fluid is followed by the kinetic
theory of resonance gases,
the switching temperature
from a hydrodynamic description to a kinetic one
is taken as $T_{\mathrm{sw}}=169$ MeV.
Ideal hydrodynamics with $T^{\mathrm{th}}=100$ MeV which is so chosen
to generate enough radial flow gives
a trapezoidal shape of $v_2(\eta)$ \cite{Hirano:2002ds,Hirano:2001eu}.
A large deviation between data \cite{Back:2002gz}
and the ideal hydrodynamic result
is seen especially in forward/backward rapidity regions.
However, just after hadronization ($T^{\mathrm{th}}=169$ MeV),
an ideal sQGP fluid gives
a triangle shape
as shown by the dotted line in Fig.~\ref{fig:v2eta}. 
(Note that the contribution from resonance decays
dilutes slightly the elliptic flow
generated in the QGP phase \cite{Hirano:2000eu}
but that the shape is not so changed.)
This means the large $v_2$ in forward/backward
rapidity regions in ideal hydrodynamic simulations 
with $T^{\mathrm{th}} = 100$ MeV is mainly 
generated in the perfect fluid of hadron phase.
When hadronic rescattering effects are taken through
the hadronic cascade model
instead of perfect fluid description of the hadron phase,
$v_2$ is not so generated in the forward region due to the
dissipation and, eventually, is consistent with
the data. 
So the perfect fluid sQGP core and the dissipative hadronic
corona picture works well also in the forward region \cite{Grassi}.
Note that the eccentricity in the CGC initial condition
is found to be 
larger than that in conventional participant scaling profiles.
This is crucial to reproduce the $v_2$ data in central
collisions (3-15\%) in our approach.
For results from
conventional parametrization based on Glauber approaches
for the initial conditions,
the deviation between ideal hydrodynamics  \cite{Huovinen:2003fa}
and the data \cite{Ackermann:2000tr,Adcox:2002ms,Back:2002gz,Ito}
which was seen previously in peripheral collisions
is interpreted only by the late hadronic viscosity \cite{HHKLN}.
On the other hand, the hybrid approach with CGC initial
conditions overpredicts the $v_2$ data
unless the early QGP viscosity
is considered in peripheral collisions.
This suggests that 
one needs a much better understanding of
initial conditions \cite{instability}
to draw the information about
the transport properties of the sQGP from the elliptic flow data.

\begin{figure}[tb]
\insertplot{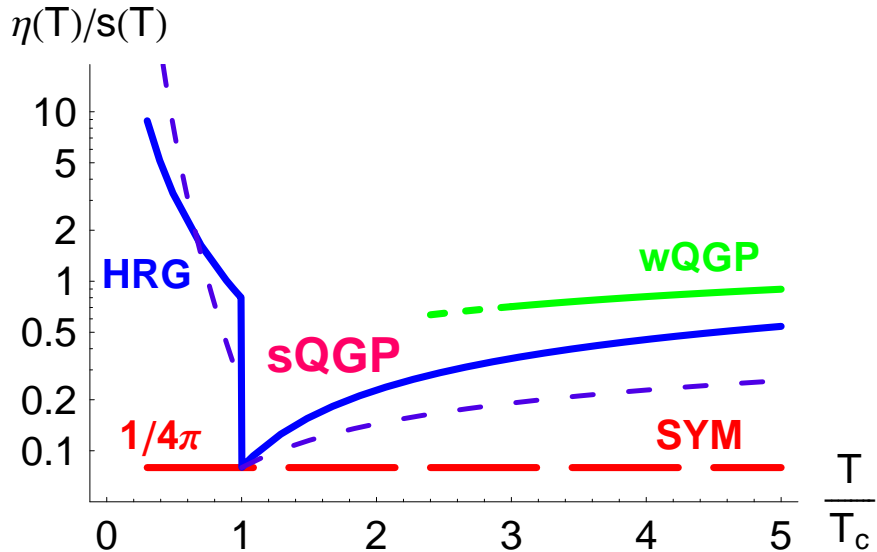}
\caption[]{Illustration of the rapid variation of the 
\textit{dimensionless ratio}
of the shear viscosity, $\eta(T)$, to the entropy density, $s(T)$.
Solid and Short dashed lines above $T_c$ 
asymptotically merge wQGP result
at high temperature region. We introduce a parameter which controls
how fast the sQGP result reaches the wQGP one.
Solid and short lines below $T_c$ corresponds to
sound velocity
in the resonance gas model $c_s^2 = 1/3$ and 1/6 respectively. 
Long dashed line is $\eta/s = 1/4\pi$ from SYM \cite{Son}.
}
\label{fig:etaovers}
\end{figure}

\section{Summary: What have we learned?}\label{concl}
At least in central collisions (up to $\sim$ 20\%),
we can establish a new picture of space-time evolution
of produced matter,
namely ``perfect fluid sQGP core
and dissipative hadronic corona",
from 
a careful comparison of hydrodynamic
results with experimental data
observed at RHIC.
Note that, in semi-central to peripheral collisions, there is a room 
for existence of additional dissipative effects in the QGP phase
from a recent quantitative analysis \cite{HHKLN} of $v_2$ data.
This comes from presently unknown details of
the initial stage in relativistic heavy ion collisions.
Further theoretical efforts for understanding
of the initial stage \cite{instability} are highly needed
to obtain a more conclusive picture.
In any case, a possibility of the picture,
``perfect fluid sQGP core
and dissipative hadronic corona",
is intriguing.

What is the physics behind this picture?
$\eta/s$ is known to be a good dimensionless measure
(in natural unit $\hbar=k_{B}=c=1$) to see
the effect of viscosity, where $\eta$
is the shear viscosity and $s$ is the entropy density.
Figures 3 and 4 show possible scenarios
for temperature dependence
of $\eta$ and $\eta/s$ deduced from the
discussion in the previous sections.
$\eta/s$ must be small in the QGP phase,
which might be comparable 
with a value $1/4\pi$
conjectured as the minimum one among any kinds of fluids \cite{Son},
and the perfect fluid assumption
can be valid. While $\eta/s$ becomes huge
in the hadron phase and the dissipation
cannot be neglected.
Shear viscosities of both phases 
are found to give $\eta \sim 0.1$ GeV/fm$^2$
around $T_c$ \cite{hg}.
So shear viscosity itself is expected to increase with temperature
monotonically.
The ``perfect fluid'' property of the sQGP is thus not
due to a sudden reduction of the viscosity
at the critical temperature $T_c$, but to
a sudden increase of the entropy density 
of QCD matter and is therefore an
important signature of deconfinement.

\section*{Acknowledgment(s)}

This work was supported in part by the United States
Department of Energy under Grant No.~DE-FG02-93ER40764.
The author would like to thank U.~Heinz, D.~Kharzeev,
M.~Gyulassy, R.~Lacey and Y.~Nara
for collaboration and fruitful discussion.

\section*{Notes}
\begin{notes}
\item[a]
E-mail: hirano@phys.columbia.edu
\end{notes}

\vfill\eject

\begin{thebibliography}{99}
\bibitem{instability}
One can find many papers
in these proceedings
for recent progress to understand
non-equilibrium aspects of gauge theories and
an initial prethermalisation 
stage in relativistic heavy ion collisions.
\end{thebibliography}
\end{document}